\documentclass[twocolumn,prl,aps,floatfix]{revtex4}
\usepackage{psfig}
\newcommand{\beq}{\begin{equation}}
\newcommand{\eeq}{\end{equation}}
\newcommand{\beqn}{\begin{eqnarray}}
\newcommand{\eeqn}{\end{eqnarray}}

\def\la{\langle} 
\def\ra{\rangle}

\def\GeV{\,\mbox{GeV}}

\begin{document}
\title{Nuclear Broadening of Transverse Momentum in Drell-Yan Reactions}
\author{
M.B.~Johnson$^a$,
B.Z.~Kopeliovich$^{b,c,d}$,
M.J.~Leitch$^{a}$,
P.L.~McGaughey$^{a}$,
J.M.~ Moss$^{a}$\\
I.K.~Potashnikova$^{b,c}$,
Ivan~Schmidt$^{b}$}
\affiliation{$^a$ Los Alamos National Laboratory, MS H846, Los Alamos,
NM 87545, USA\\
$^b$ Departamento de F\'isica, Universidad Tecnica Federico Santa
Mar\'ia, Valpara\'iso, Chile\\
$^c$ Joint Institute for Nuclear Research, Dubna, Russia\\
$^d$ Institute of Theoretical Physics, Heidelberg University, 
Heidelberg, Germany}


\begin{abstract}
 Data for Drell-Yan (DY) processes on nuclei are currently available from
fixed target experiments up to the highest energy of $\sqrt{s}=40\GeV$.
The bulk of the data cover the range of short coherence length, where the
amplitudes of the DY reaction on different nucleons do not interfere. In
this regime, DY processes provide direct information about broadening of
the transverse momentum of the projectile parton experiencing
initial-state multiple interactions. We revise a previous analysis of data
from the E772 experiment and perform a new analysis of broadening
including data from the E866 experiment at Fermilab. We conclude that the
observed broadening is about twice as large as the one found previously.
This helps to settle controversies that arose from a comparison of the
original determination of broadening with data from other experiments and
reactions.

\smallskip

\end{abstract}
\maketitle    

\vspace*{0.5cm}

\section{Introduction}

Understanding how partons propagate in a nucleus reveals precious
information about the early stages of hadronization that is difficult to
obtain by other means. This understanding also plays a role in using
partons as a sensitive probe of matter produced in relativistic heavy ion
collisions. One of the important observables is broadening of the
transverse momentum of produced particles due to multiple interactions of
a parton in the medium. This effect can be studied for the nuclear medium
in Drell-Yan reactions \cite{na10,e772,mmp} and deep-inelastic scattering
\cite{hermes,jlab} on nuclear targets. The former is an especially clean
tool, since the produced dilepton, which has no final state interactions,
carries undisturbed information on the transverse momentum of the parton
initiating the reaction.

A parton propagating through a nuclear medium experiences Brownian motion
in the transverse momentum plane \cite{pirner}, with the increase of the
mean transverse momentum squared being proportional to the length of the
path times the medium density $\rho(z)$,
 \beq
\Delta\la p_T^2(L)\ra = 2C(s)\int\limits_0^L dz\,\rho(z),
\label{10}
 \eeq 
 where $z$ is the longitudinal coordinate. The energy dependent
dimensionless factor $C$ characterizes multiple interactions of the parton
with bound nucleons. In the light-cone dipole description of the
broadening process \cite{dhk,jkt,at}) $C$ is related to the small
separation behavior of the dipole-nucleon cross section,
 \beq
C(s)=\left.\frac{d\sigma_{\bar qq}(r_T,s)}
{dr_T^2}\right|_{r_T=0}\ .
\label{20}
 \eeq
 The cross section $\sigma_{\bar qq}(r_T,s)$ introduced in \cite{zkl}
varies with transverse $\bar qq$ separation $\vec r_T$ and energy. This
quantity is difficult to calculate, since it involves nonperturbative
effects, but it can be fitted to data for the photoabsorption cross
section and the proton structure function $F_2(Q^2,x)$, which probe a wide
variety of separations and energies \cite{gbw,kst2}.  Since each of the
scatterings is accompanied by a loss of energy, there is a close
connection between the multiple scattering that leads to momentum
broadening and initial state gluonic energy loss \cite{bdmps,vitev}.
It was found, however, that this induced energy loss effects the DY 
process much less than the vacuum gluon radiation initiated by the 
interaction on the nuclear surface \cite{e-loss}.

In the parton model, a single interaction of a quark propagating through
the medium can be viewed as a quark-gluon correlation in a nucleus in its
infinite momentum frame. The corresponding twist-4 term \cite{qs} cannot
be calculated, but is rather fitted to data. This phenomenology has not
been successful so far, with the parameter extracted from data varying by
a factor of 20, depending on assumptions \cite{guo} and also demonstrating
a strong process dependence \cite{ww}.

The first observation of nuclear enhancement of large transverse momenta
was made in hadron-nucleus collisions \cite{cronin}. A sensitive tool to
measure the transverse momentum broadening of quarks propagating through
nuclei is heavy dilepton production in the Drell-Yan reaction. The leptons
have no interaction in the nuclear medium and thus carry undisturbed
information about the transverse momentum of the quark. Nuclear broadening
of the mean transverse momentum squared of the dilepton is defined as
 \beq
\Delta\la p_T^2\ra = \la p_T^2\ra_A -
\la p_T^2\ra_N\ .
\label{22}
 \eeq
 Perturbative QCD predicts $d\sigma_{DY}/dp_T^2\propto 1/p_T^4$ (for
transversely polarized virtual photons) at large $p_T$. Therefore, the
mean value $\la p_T^2\ra$ is divergent, and the broadening of
Eq.~(\ref{22}) might be ill defined (see discussion in \cite{krtj}). Of
course, the experiment imposes a cut off.  Thus, the experimentally
observed broadening is finite, but it may depend on the experimental
acceptance.  This divergence, however, sometimes cancels in the broadening
for several reasons. First of all, perturbative QCD predicts no nuclear
effects at large $p_T$, i.e., $d\sigma_{DY}(pA)/dp_T^2 \to
A\,d\sigma_{DY}(pp)/dp_T^2$ (see, e.g., \cite{knst}). Secondly, nuclear
shadowing is expected to vanish at large $p_T$, especially at an energy as
low as $800\GeV$, the energy of our current analysis. In this case, the
divergent high-$p_T$ part of the integrations cancel in (\ref{22}), and
one can measure the nuclear broadening effect within a restricted interval
of $p_T$ where the data exhibit nuclear effects.

\begin{figure}[tbh]
\centerline{\psfig{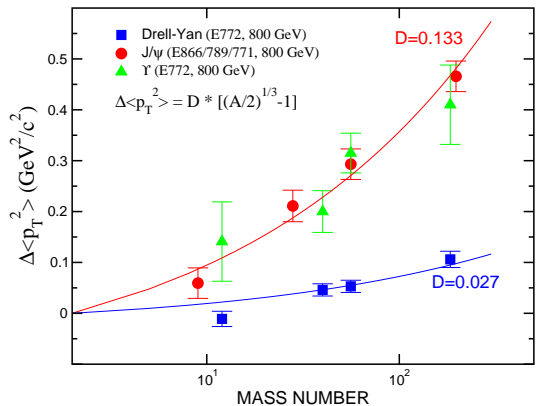}}
  \protect\caption{Broadening of transverse momentum in production of
heavy dileptons (squares), $J/\Psi$ (circles) and $\Upsilon$ (triangles)
in $pA$ collisions at $800\GeV$. The data are from the E772, E789, and
E866 experiments at Fermilab \cite{e772,e772a,e789,e866-pt}. The curves
are results of a fit with the parametrization $\Delta\la p_T^2\ra
=D\left[(A/2)^{1/3}-1\right]$.}
 \label{e772-old}
 \end{figure}

An analysis of broadening in the Drell-Yan reaction was presented in
\cite{mmp} using data from the E772 experiment at Fermilab. The results of
the analysis for $\Delta\la p_T^2\ra_{DY}$ are depicted in
Fig.~\ref{e772-old} as function of $A$. \\[0.2cm]
 
Following Eq.~(\ref{10}), one can parametrize the $A$-dependence as 
 \beq
\Delta\la p_T^2\ra =D\left[(A/2)^{1/3}-1\right]
 \label{25}
 \eeq
 A fit to the data points for the Drell-Yan reaction in Fig.~\ref{e772-old}
results in $D=0.027\GeV^2$, which corresponds to $C=2$ in (\ref{10}).

\begin{figure}[tbh]
\centerline{\psfig{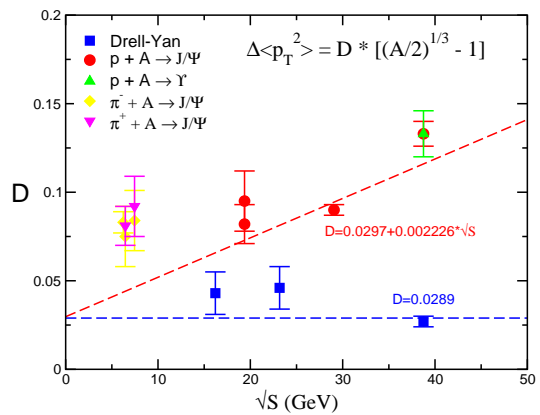}}
  \protect\caption{Energy dependence of the parameter $D$ in (\ref{25})
fitted to data for dilepton and heavy quarkonium production in $pA$ and
$\pi A$ collisions.}
 \label{e-dep}
 \end{figure}

The same figure presents nuclear broadening in $J/\Psi$ and $\Upsilon$
production at the same energy. Comparing with this data and results
available at lower energies, one can make the following observations: (i)
the magnitude of broadening in reactions of heavy dilepton and quarkonium
production differ by a factor of 5, while one would expect the Casimir
factor $9/4$. Indeed, heavy quarkonia should be produced via gluons, which
interact in the medium more strongly than quarks. (ii) The errors for
Drell-Yan data are small compared to $J/\Psi$ production, whereas the
latter is measured with much higher statistics. (iii) Comparison with
broadening measured at lower energies presented in Fig.~\ref{e-dep}
suggests that broadening in heavy quarkonium production should rise with
energy, whereas it is seen to be rather flat, or even perhaps drop, at
high energies in the Drell-Yan reaction.  \\[0.2cm]

The latter observation is difficult to explain, since the cross section
controlling the multiple interactions in the medium rises with energy.
(iv) The dipole approach, Eq.~(\ref{20}), based on a phenomenological color dipole
cross section fitted to data, predicts the factor $C$ \cite{jkt} to be
about twice as large as one suggested by the data presented in
Fig.~\ref{e772-old}. \\

In this paper we revise the result of the previous analysis \cite{mmp} of
the E772 data for Drell-Yan nuclear ratios, combining them with the
brand-new data \cite{webb} from the E866 experiment for Drell-Yan cross
section in $pD$ collisions.  The improvement in the analysis arises from
the fact that the E866 experiment provides more accurate $pD$ cross
section data at small $p_T$ than that available from E772. Additionally,
it was assumed with no good justification that the same simple
parametrization of the $p_T$-dependence is valid for both $pD$ and $pA$
collisions, which dictated a specific shape for the ratios and led to a
substantial underestimate of the errors for $p_T$-broadening. Our new
analysis is less model-biased, but it results in somewhat larger errors.  
We also perform here an analysis of new data from the E866 experiment,
which leads to a broadening about twice as large as in \cite{mmp},
although with larger experimental errors. This result resolves the
problems stated earlier.

\section{Light-cone dipole description of nuclear broadening}

When a projectile scatters from a nucleus, one can treat the target nucleons as independent with some degree of accuracy if viewed in the rest frame of the nucleus, since the bound nucleons are sufficiently well separated in that frame.
However, when boosted to the infinite momentum frame the same reaction looks different since the nucleons may start 
communicating via their parton clouds at small Bjorken $x$. Indeed the 
Lorentz contraction is proportional to $1/x$, therefore it is much 
stronger at large, than at small $x$. As a result, the parton clouds are 
still separated for $x>x_c$, but overlap and interact at smaller $x$, 
$x<x_c$. The border line $x_c$ can be estimated as,
 \beq
x_c \sim \frac{1}{m_N\,R_A}\ ,
\label{26}
 \eeq
 where $R_A$ is the nuclear radius. Note that this is only an order of
magnitude estimate. There is a numerical factor in (\ref{26}) that varies
considerably for different species of partons and the way the
parton distribution is probed \cite{krt2}.

The interaction between partons at $x<x_c$ leads to a modification of
their transverse momentum distribution: suppression of small momenta,
enhancement at intermediate, and no effect at large transverse momenta.
This phenomenon is known as the saturation effect \cite{glr}, or color
glass condensate \cite{mv}.  One should keep in mind, however, that onset
of this effect happens only at $x<x_c$.

 The partonic interpretation of hard reactions at small Bjorken $x$ is
known to be Lorentz non-invariant. Deep-inelastic scattering (DIS) looks
like absorption of the virtual photon by a quark (antiquark) that belongs
to the target in the infinite momentum frame of the latter. The reaction looks,
however, quite different in the target rest frame, as the interaction with a
$\bar qq$ fluctuation of the high-energy virtual photon. Of course, all
observables, including the structure function $F_2(x,Q^2)$, which is
proportional to the total photon-target cross section, are Lorentz
invariant.

The Drell-Yan reaction is usually interpreted as annihilation of a quark
with an antiquark belonging to the beam and target respectively (or vice
versa), $\bar qq \to \bar ll$ \cite{dy}. While this is correct in the rest
frame of the dilepton, this interpretation of a heavy dilepton production
is not applicable in target rest frame. Indeed, in order to satisfy the
kinematics annihilating with the projectile, the target parton must move
with momentum $p_L\sim p_T/x_2$, where $x_2$ in standard notation is
the Bjorken $x$ of the target parton. At small $x_2$, $x_2\ll1$, this would mean
presence of highly energetic partons in the stationary target. The proper
space-time interpretation in this case is analogous to the
Weitz\"acker-Williams mechanism of electromagnetic bremsstrahlung. Namely,
a beam quark (antiquark) develops a fluctuation, $q\to q\gamma^*$ (with
$\gamma^*\to\bar ll$), which interacts with the target freeing the
dilepton\cite{k,kst1}.

What looked like overlap of parton clouds at small $x_2$ in the infinite
momentum frame, now takes the form of interference between the amplitudes
of a parton interacting with different nucleons. On the same footing it can
also be interpreted as the lifetime of the $|q\gamma^*\ra$ fluctuation,
given by the uncertainty principle
 \beq
 l_c=\frac{1}{q_L} = 
\frac{2E_q}{M^2_{q\gamma^*}-m_q^2}\ , \label{28}
 \eeq
 where $q_L$ is the longitudinal momentum transfer in the $q\to q\gamma^*$
fluctuation.  The effective mass squared of the fluctuation is
$M_{q\gamma^*}^2=M_{\gamma^*}^2/\alpha+m_q^2/(1-\alpha)+k_T^2/\alpha(1-\alpha)$,
where $\alpha$ is the fraction of the light-cone momentum of the parent
quark taken by the dilepton ($\gamma^*$), and $k_T$ is the $\gamma^*-q$
relative transverse momentum.  Hereafter we do not discriminate between
coherence time and length, assuming that the quark has the speed of light.
Evidently, the same length $\sim1/q_L$ defines the maximal longitudinal
distance between two scattering centers that can interfere. This is why
the distance, Eq.~(\ref{28}), is also called coherence length.

In the case of a nuclear target, the coherence length is of great importance.
If this time interval is small compared to the mean spacing of nucleons in
the nucleus, dileptons are radiated in the Bethe-Heitler regime, i.e., with
no interference between the amplitudes of radiation from different
nucleons. At the another extreme, $l_c\gg R_A$, the Landau-Pomeranchuk
phenomenon governs the radiation. Namely, interferences suppress radiation
with small transverse momenta, enhance it at intermediate values of few
GeV, and make no changes at higher momentum transfer. In both cases, the
linear dependence on the path length, or nuclear thickness,
Eq.~(\ref{10}), holds, although with different factors $C$ \cite{krtj}.
The borderline between the two regimes, $l_c=R_A$, corresponds to the
relation for $x_c$, Eq.~(\ref{26}).

\subsection{Short coherence length}

The mechanism of nuclear broadening depends on how long the coherence
length is compared to the nuclear size. In the regime of short coherence
length, $l_c\ll R_A$, the dilepton fluctuation appears for a very short
time deep inside the nucleus and is immediately released on mass shell by
the interaction, which is assumed to be the same as on a free nucleon.
Therefore, this interaction does not lead to any change in the transverse
momentum dependence, i.e., it provides no broadening. However, the initial
state interactions of the projectile quark affect the transverse momentum
distribution of the quark, which acquires the shape \cite{jkt}
 \beqn
\frac{dN_q}{d^2p_T} &=&
\int d^2r_1\,d^2 r_2\,e^{
i\,\vec p_T\,(\vec r_1 - \vec r_2)}\,
\Omega^q_{in}(\vec r_1,\vec r_2)\nonumber\\
&\times& e^{-{1\over2}\,
\sigma^N_{\bar qq}(\vec r_1-\vec r_2,x)\,T_A(b)}\ .
\label{30}
 \eeqn
 Here $\Omega^q_{in}(\vec r_1,\vec r_2)$ is the density matrix describing
the impact parameter distribution of the quark in the incident hadron.
We assume a Gaussian shape,
 \beq
\Omega^q_{in}(\vec r_1,\vec r_2) =
\frac{\la p_0^2\ra}{\pi}\,
e^{-{1\over2}(r_1^2+r_2^2)
\la p_0^2\ra}\ ,
\label{33}
 \eeq
 where $\la p_0^2\ra$ is the mean value of the primordial
transverse momentum squared of the quark.

The transverse momentum of the radiated dilepton reflects the broadening acquired by the quark in the nuclear medium.  However, the effect is weaker than for the quark, since the
transverse momenta of the dilepton and quark are connected by the
relation,
 \beq
q_T^{\bar ll} = \alpha\,p_T^q\ ,
\label{35}
 \eeq
 where $\alpha < 1$ is the fractional momentum of the $\bar ll$. This
relation clearly demonstrates that the effect of broadening cannot be
translated into a modification of the quark distribution function (as is
frequently done), in the regime of short coherence length. Thus, it
confirms the breakdown of $k_T$-factorization imposed by the initial state
interaction, as suggested in \cite{bbl}.

\subsection{\bf No-shadowing sum rule}

\begin{figure}[tbh]
\centerline{\psfig{figure=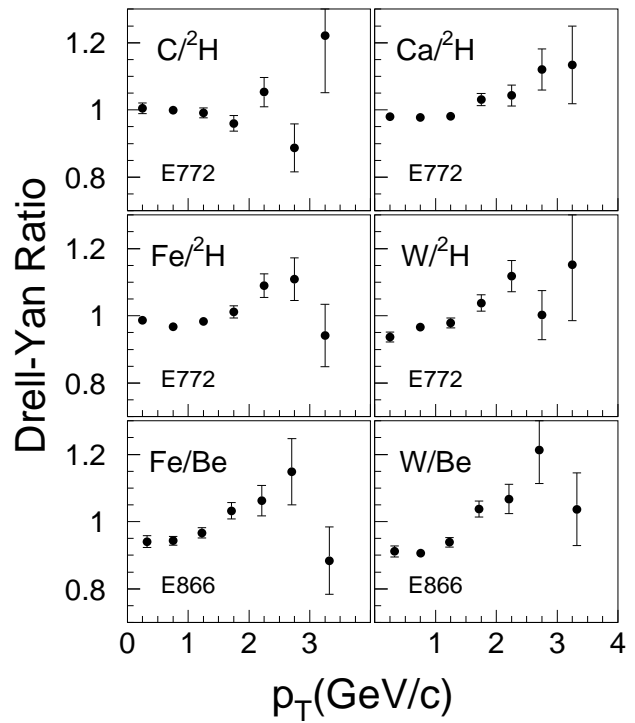,width=8cm}}
  \protect\caption{The nucleus-to-deuterium and -beryllium ratios of
dilepton production cross sections as function of transverse momentum. The
data are from the E772 \cite{e772} and E866 \cite{e866-pt} experiments  at
$800\GeV$.}
 \label{pt-e772}
 \end{figure}

To the extent that the coherence length for the Drell-Yan reaction is
short compared to the internucleon spacing in the nucleus, no shadowing is
possible. This means that in spite of the nuclear modification of the
transverse momentum dependence of Drell-Yan cross section, the number of
quarks is conserved. This fact can be represented in the form of a sum
rule,
 \beq
\int d^2p_T\,\frac{d\sigma_{DY}(pA)}{d^2p_T} =
A\,\int d^2p_T\,\frac{d\sigma_{DY}(pN)}{d^2p_T}\ .
\label{40}
 \eeq
 The nuclear ratios measured in the E772 and E866 experiments at $800GeV$
depicted in Fig.~\ref{pt-e772} have a tendency to satisfy this constraint.
Indeed, the ratios are below one at small $p_T$, but rise above one at 
larger momentum transfers.
 The weak nuclear suppression observed at small $p_T$, where the absolute
cross section is maximal, seems to be compensated by a stronger
enhancement at larger $p_T$ in accordance with the sum rule
Eq.~(\ref{40}). Data in the small $p_T < 1-2\GeV$ domain, where the main
part of the cross section comes from, are most important for broadening.
In this region the E866 data in Fig. 3 demonstrate stronger nuclear
suppression than the E772 data. The observed deviation from unity at small
$p_T$, which is responsible for broadening, is about twice as large in
E866 as in E772 data.

Notice that the $p_T$-dependence of the ratios in Fig.~\ref{pt-e772} has
the typical shape predicted for the regime of saturation (of quarks), or
color glass condensate \cite{mv}. Therefore, it might be tempting to
interpret such data as an observation of this phenomenon. We warn again
that this would be a wrong conclusion, since coherence and saturation are
impossible in the short coherence length regime; initial state interaction
effects imitate saturation.

\subsection{Long coherence length} 

In this regime, $l_c\gg R_A$, the amplitudes $\gamma^*\to\bar ll$ for
radiating a heavy photon from different nucleons interfere with each
other.  This interference, which modifies the $p_T$ distribution of
radiation, is known as the Landau-Pomeranchuk effect and is a genuine
effect of saturation.

Alternatively, one may say that a long-lived fluctuation $q\to\gamma^*q$
that propagates through the entire nucleus may be freed as a result of the
multiple interactions, in contrast to what happens in the regime of short
$l_c$, where the fluctuation is freed after an interaction with a single
nucleon.  These multiple interactions also lead to a different type of
broadening mechanism. Its origin can be understood from the fact that most
of the $q\to\gamma^*q$ fluctuations in the quark are invisible in the
sense that their interactions with the target are so weak that the target
has restricted resolving power or ability to discriminate a hard
$q\gamma^*$ fluctuation, with large intrinsic transverse momentum $k_T$,
from a bare quark.  However, because of the larger momentum transfer
allowed by the multiple interactions in the nuclear medium, the resolution
provided by the nucleus is greater than that of a free nucleon.  The
nucleus can therefore free harder fluctuations than a free nucleon. This
is the source of nuclear broadening in the large coherence length regime.

The same effect can be interpreted within the color dipole approach in
terms of color filtering. The nucleus filters out projectile dipoles of
smaller size than is possible on a free nucleon, leading to larger
transverse momenta. In this case, instead of Eq.~(\ref{30}), the
transverse momentum distribution of dileptons radiated by a quark
interacting with a nucleus is given by the factorized dipole formula
\cite{kst1},
 \beqn
&& \frac{d\sigma(qA\to \bar ll X)}
{d^2q_T} =
\int d^2b\int d^2r_1d^2r_2\,
e^{i\vec q_T(\vec r_1-\vec r_2)}\,
\nonumber\\ &\times&
\Psi_{q\gamma^*}^\dagger(\vec r_1,\alpha)\,
\Psi_{q\gamma^*}(\vec r_2,\alpha)
\left[1 - e^{-{1\over2}\sigma^N_{\bar qq}(r_1,x_2)T_A(b)}
\right.\nonumber\\ &-& \left.
e^{-{1\over2}\sigma^N_{\bar qq}(r_2,x_2)T_A(b)} +
e^{-{1\over2}\sigma^N_{\bar qq}(\vec r_1-\vec r_2,x_2)T_A(b)}
\right]\ .
\label{80}
 \eeqn
 Here $\Psi_{q\gamma^*}(\vec r,\alpha)$ is the pQCD calculated light-cone
distribution amplitude, which describes the dependence of the
$|q\gamma^*\ra$ Fock component on transverse separation and fractional
momentum. The nuclear thickness function at impact parameter $b$ is given
by the integral of the nuclear density along the quark trajectory,
$T_A(b)=\int_{-\infty}^\infty dz\,\rho_A(b,z)$.

The ratio of the nuclear to nucleon cross sections of the Drell-Yan
reaction calculated in \cite{kst1} has a shape typical for the Cronin
effect. Such a modification of the transverse momentum dependence is a
genuine effect of saturation (or its onset). In spite of the similarity
with the analogous results in the short $l_c$ regime, the mechanisms and
their formulations in Eqs.~(\ref{30})-(\ref{80}) are different.
Additionally, the magnitude of the Cronin effect in the long $l_c$ regime
is smaller \cite{krtj}.

\section{Analysis of E772/866 data for Drell-Yan reaction}

Here we determine the nuclear broadening $\Delta\la p_T^2\ra$ of dileptons
produced on different nuclear targets relying on data for $p_T$-dependent
ratios depicted in Fig.~\ref{pt-e772}. These include the
nucleus-to-deuterium ratios $R^{A/D}$ from the E772 \cite{e772}
experiment, and nucleus-to-beryllium ratios $R^{A/Be}$ from the E866
experiment \cite{e866-pt}, both at $800\GeV$. These results are quite
reliable since they are least affected by systematic uncertainties. The
ratios $R_i$ are available in 7 bins in $p_T$, $i=1-7$.  Correspondingly,
one can calculate the broadening, Eq.~(\ref{22}), as
 \beq
\Delta\la p_T^2\ra = 
\frac{\sum_i \sigma_i\,R_i\,p_i^2}{\sum_i \sigma_i\,R_i}
- \frac{\sum_i \sigma_i\,p_i^2}{\sum_i \sigma_i}\ ,
\label{60}
 \eeq 
 where for each $i$-th bin
 \beq
\sigma_i=\int\limits_{(p_T)_i^{min}}^{(p_T)_i^{max}}
d^2p_T\,\frac{d\sigma_{DY}^N}{d^2p_T}\ ,
 \label{70}
 \eeq
 with $p_i^2$ the mean value of $p_T^2$ within the $i$-th bin. In addition
to the seven $p_T$-intervals in data depicted in Fig.\ref{pt-e772}, we
have included an eighth interval in Eq.~(\ref{60}) covering
$(p_T)_7^{max}<p_T<\infty$ and assumed that $R_8=1$ in accordance with
perturbative QCD expectations \cite{knst} (see, however, next section).
Since the cross section falls steeply at large $p_T$, this point should
not substantially affect the results.

The differential cross section for the Drell-Yan reaction on a free
nucleon, $d\sigma_{DY}^N/d^2p_T$, is more affected by systematic
uncertainties than the ratio is.  We rely on the recently published
results of the E866 experiment for the differential cross section in $pp$
and $pD$ collisions \cite{webb}. We use this data in (\ref{70}) for the
analyses of all data presented in Fig.~\ref{pt-e772}.

We parametrize the differential cross section of the Drell-Yan reaction as
 \beq
\frac{d\sigma}{dp_T^2} = N\,
\frac{\left(1+\frac{p_T^2}{\lambda_1^2}\,
e^{-p_T^2/\lambda_2^2}\right)}
{(1+p_T^2/\lambda_3^2)^n}\ .
\label{90}
 \eeq
 The factor in the numerator is introduced to describe a possible forward 
minimum in the cross section indicated by data \cite{webb}.

\begin{figure}[tbh]
\centerline{\psfig{figure=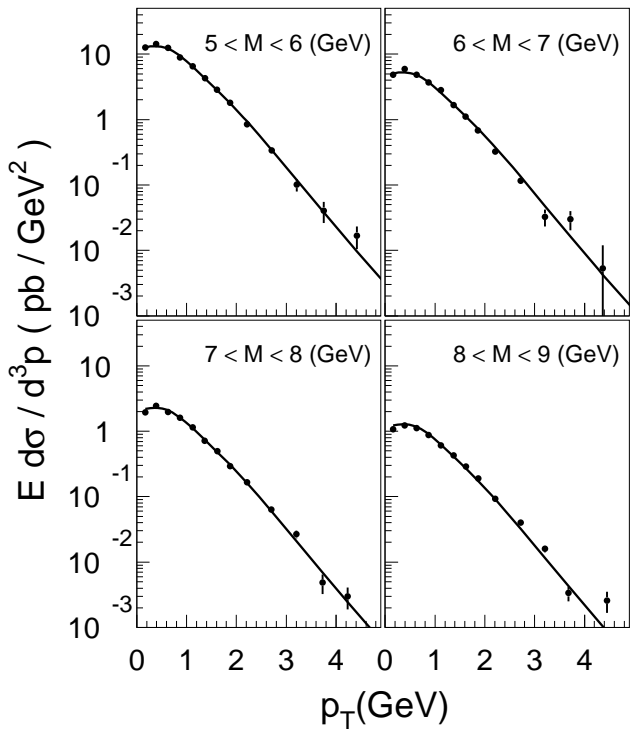,width=8.3cm}}
  \protect\caption{Differential cross section of Drell-Yan reaction in
proton-deuteron collisions. Data are from the E866 experiment
\cite{e866-pt} at $800\GeV$ for different intervals of dilepton effective
mass. Curves show the result of the global fit with common parameters,
Eq.~(\ref{90}),  controlling the shape of the cross section.}
 \label{fit}
 \end{figure}

We found that the shape of the $p_T$ dependence does not vary with
dilepton mass and $x_2$, as one can see in Fig.~\ref{fit}. 
 A global fit to Drell-Yan data from $pD$ collisions measured by the E866
experiment for different $M$ and $x_2$ bins \cite{webb} with common
parameters $n$, $\lambda_i$ and different normalization factors $N$, led
to quite a good description, $\chi^2 = 83.3$ with $N_{d.o.f.}=52$. The
values of the parameters are $\lambda_1=0.78 \pm 0.08$; $\lambda_2=0.65
\pm 0.04$; $\lambda_3=3.31 \pm 0.21$;  $n=7.00\pm0.66$. The results of the
fit and the data are depicted in Fig.~\ref{fit}. The error bars include
both statistical and systematic errors added in quadrature.

 Note that at large $p_T$ the $pD$ cross section in Eq.~(\ref{90}) is
falling as $p_T^{-14}$, much steeper than one may expect from perturbative
QCD. This happens due to restrictions imposed by finiteness of energy,
namely, both $x_1$and $x_2$ rise with $p_T$ towards the kinematic limit
causing an additional suppression of the cross section \cite{krtj}.
Additionally, available data cover a rather restricted interval of $p_T <
4\GeV$.

\begin{figure}[tbh]
\centerline{\psfig{figure=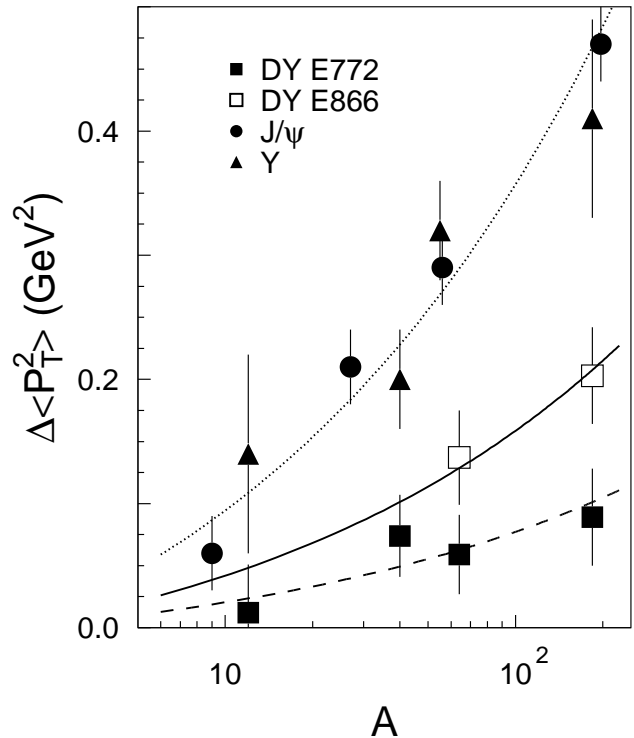,width=8cm}}
  \protect\caption{Results of the present analysis for broadening in the
Drell-Yan reaction on different nuclei. Closed squares correspond to E772
data for $C,\ Ca,\ Fe$ and $W$. Open squares correspond to E866 data for
$Fe$ and $W$. For comparison, broadening for $J/\Psi$ and $\Upsilon$, the
same as in Fig.~\ref{e772-old}, are also shown.}
 \label{results}
 \end{figure}

 With the cross section in Eq.~(\ref{90}) we performed an analysis of data
from both experiments, E772 and E866.  While the ratios measured in E772
are free from major systematic uncertainties, the absolute cross sections
are substantially less reliable. Indeed, comparison of the differential
cross sections of the Drell-Yan process measured for $p-D$ collisions in
the two experiments reveals substantial differences. Therefore, we rely on
the $p-D$ cross section from the E866 experiment. The results of such a
combined analysis of the E772 ratios are depicted in Fig.~\ref{results} by
closed squares.

The ratios measured in the E866 experiment are for a nucleus relative to
beryllium. We use the chain relation between the two ratios,
 \beq
R_{A/D}=R_{A/Be}\,R_{Be/D}\ ;
\label{99}
 \eeq
 however, the beryllium-to-deuterium ratio is unfortunately lacking.  
Nevertheless, this cannot be a significant correction, since beryllium is
a light nucleus.  Indeed, the carbon to deuterium ratio depicted in
Fig.~\ref{pt-e772} is compatible with no nuclear effects at all.  If we
simply fix $R_{Be/D}=1$, the broadening given in Eq.~(\ref{60}) will be
underestimated. However, one can do better by relying on a theoretically
predicted value for $R_{Be/D}$ \cite{mikk} with Eq.~(\ref{30}). Of course,
this introduces an uncertainty related to model dependence, but this
uncertainty should not exceed few percent of the observed broadening. The
results of this analysis of the E866 data are depicted in
Fig.~\ref{results} by empty squares.
 
Notice that the cross section ratios are affected by systematic
uncertaities much less than the absolute cross sections. The ratios
measured in both experiments have mainly an overall systematic uncertainty
$1\%$ in E866 \cite{e866-pt}, and $2\%$ in E772 \cite{e772} measurements.  
One can see from Eq.~(\ref{60}) that an overall variation of all $R_i$
leaves the broadening unchanged. Therefore we could disregard the
systematic errors of the ratios.

 If we fit the data in Fig.~\ref{results} assuming a linear dependence on
$A^{1/3}$, in accordance with (\ref{25}), and do this separately for the
E772 and E866 data, we obtain,
 \beqn
D(E772)=(0.029 \pm 0.008)\,(\GeV/c)^2\ ;
\nonumber\\
D(E866)=(0.059 \pm 0.009)\,(\GeV/c)^2\ .
\label{100}
 \eeqn

 The value of $D(E772)$ is close to the result of previous analysis
\cite{mmp} depicted in Fig.~\ref{e772-old}, although with a considerably
larger error. However, the E866 data suggest a much stronger broadening.
Nevertheless, one may consider the results Eq.~(\ref{100}) as nearly
consistent since they lie within two standard deviations.  If the
broadening according to the E866 data were actually about twice as large
as that given by the previous analysis \cite{mmp} of the E772 data, this
would settle the contradiction with theoretical expectations and
conclusions drawn from the data for the broadening of $J/\Psi$ and
$\Upsilon$ mentioned above.

\section{Beyond the data}

Although QCD predicts no nuclear effects at high $p_T$, the data in
Fig.~\ref{pt-e772} show considerable nuclear enhancement at intermediate
values of $p_T$. This is the well-known Cronin effect first observed for
hadronic beams \cite{cronin}. It is not clear from the data how far in
$p_T$ this effect can propagate. The assumption made above for the last
unobserved $p_T$ interval, $R_8=1$, is probably incorrect, therefore the
broadening effect shown in Fig.~\ref{results} could be be underestimated.  
The ad hoc shape of $R(p_T)$ used in \cite{mmp} is difficult to justify
and does not suggest a solution to the problem.

In order to get a hint about how much of the effect might have been
missed, one may make use of a theoretically calculated value for $R_8$. We
rely on the dipole formalism, Eqs.~(\ref{30})-(\ref{35}), from which one
is able to calculate the $p_T$ dependence of the Drell-Yan reaction on
proton and nuclear targets in a parameter-free way and explain the data in
Fig.~\ref{pt-e772} well with no adjustment of parameters. The dipole
approach also provided the first quantitative explanation of the Cronin
effect in available data for hadron-nucleus collisions and correctly
predicted the effect for RHIC \cite{knst}. Therefore we believe that a
calculated value of $R_8$ should be rather reliable. Doing this,
Eq.~(\ref{60}) leads to somewhat larger values for the broadening.
 
The fit of Eq.~(\ref{25}) to this data leads to new values of parameter $D$,
 \beqn
 \tilde D(E772)=(0.038 \pm 0.008)\,(\GeV/c)^2\ ;
\nonumber\\
 \tilde D(E866)=(0.069 \pm 0.009)\,(\GeV/c)^2\ .
\label{120}
 \eeqn

This procedure involving theoretical calculations may be considered as a
means to estimate the systematic uncertainty. The sign (positive) we
obtain for the uncertainty is certainly correct, and its magnitude is not
large. Note also that these corrections are correlated for the two sets of
data, and therefore one must apply them consistently to both the the
results of the E772 and E866 experiments.

\section{Broadening versus coherence length}

The mechanism of broadening is expected to correlate with the coherence
length \cite{jkt,knst,krtj} as described above. The coherence length of
the Drell-Yan process is controlled by the value of $x_2$, $l_c\sim 1/m_N
x_2$ \cite{e-loss}. Thus, at large $x_2$ the dilepton is produced inside
the nucleus and only half of the nuclear thickness contributes to
broadening. However, at small $x_2$ the Drell-Yan pair is radiated outside
the nucleus and the entire nuclear thickness is effective. It would be a
spectacular observation to see a manifestation of this effect in data.

\begin{figure}[tbh]
\centerline{\psfig{figure=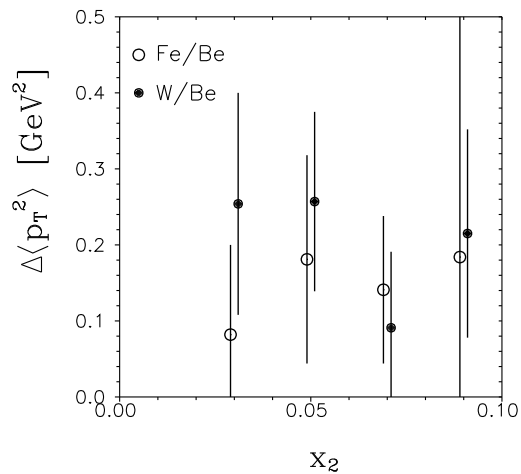,width=8cm}}
  \protect\caption{Broadening of the mean transverse momentum squared in
the Drell-Yan reaction measured \cite{e866-pt,e866-x2} on tungsten and
iron targets relative to beryllium, $\Delta\la p_T^2\ra = \la
p_T^2\ra_A-\la p_T^2\ra_{Be}$.}
 \label{x2-dep}
 \end{figure}

Using recent data from the E866 experiment for the $p_T$-dependent ratios
binned in $x_2$ \cite{e866-pt,e866-x2} we can also test the predicted
$x_2$-independence of broadening. We have performed a similar analysis of
the $Fe/Be$ and $W/Be$ ratios in different $x_2$ bins and the results are
depicted in Fig.~\ref{x2-dep}

Unfortunately, within the rather large statistical errors one cannot draw
a definite conclusion regarding a possible variation of broadening with
$x_2$.

\section{Conclusions}
 
Analysis of data from the E866 experiment for the DY reaction on nuclear
targets has resulted in a determination of the magnitude of the nuclear
broadening parameter $D(E866)$ given in (\ref{100}) and (\ref{120}), which
is about twice as large as the one suggested by the E772 data. This result
helps to resolve the controversies that were a partial motivation for this
analysis. One of these was the expectation that the observed broadening in
DY would be a factor of $4/3$ less than that observed in $J/\Psi$ and
$\Upsilon$, as dictated by the Casimir factor. The updated result is
consistent with this expectation and also with the expectation of a rising
energy dependence of broadening. And, last but not least, the value of
$D(E866)$ agrees with the parameter-free theoretical prediction of
\cite{jkt}.

This new result also agrees with broadening observed by the HERMES
experiment at HERA in hadron production in deep-inelastic scattering on
nuclei \cite{hermes}. Indeed, both measurements are well described
within the same color-dipole approach \cite{with-with}.

Since most of the data from both experiments E772 and E866 correspond to
the regime of short coherence length, the observed broadening is directly
connected via the relation in Eq.~(\ref{35}) to the broadening of a quark
propagating through the nuclear medium. In the regime of long coherence
length the mechanism of broadening changes and takes the form of color
filtering in hard coherent scatterings. To see a possible variation of
broadening as function of coherence length, we performed an analysis of
data additionally binned in $x_2$. Unfortunately, the restricted
statistics did not allow us to draw a definite conclusion from this
analysis.

Broadening of the transverse momentum of a parton in a medium leads to
induced gluon radiation, i.e., to induced energy loss. This is the basis
of the effect called jet quenching which is considered to be a sensitive
probe for properties of matter created in heavy ion collisions.

Note that broadening itself is much less dependent on models for
hadronization, than jet quenching. Therefore it is a better and more
certain probe.

\vspace{2mm}

\noindent {\bf Acknowledgments}:  This work was supported in part by U.S.
Department of Energy, the Research Ring "Center of Subatomic Studies"
(Chile), Fondecyt (Chile) grants 1030355, 1050519 and 1050589, and by DFG
(Germany) grant PI182/3-1.

\end{document}